\documentclass[twocolumn]{aastex6}
\usepackage{natbib}
\usepackage{color}

\usepackage{color}
\usepackage{hyperref} 

\def    \apjl  		{\rm {ApJL}}

\def    \apjl  		{\rm {ApJL}}

\def	\cm		{\,{\rm {cm}}}
\def	\K		{\,{\rm K}}
\def	\g		{\,{\rm {g}}}
\def	\mum	{\,{\mu \rm{m}}}

\begin{document}
\shorttitle{Relativistic gas drag}
\shortauthors{Thiem Hoang}
\title{Relativistic Gas Drag on Dust Grains and Implications}
\author{Thiem Hoang}
\affil{Korea Astronomy and Space Science Institute, Daejeon 34055, Korea; thiemhoang@kasi.re.kr}
\affil{Korea University of Science and Technology, 217 Gajeong-ro, Yuseong-gu, Daejeon, 34113, Korea}

\begin{abstract}
We study the drag force on dust grains moving at relativistic velocities through interstellar gas and explore its application. First, we derive a new analytical formula of the drag force at high energies and find that it is significantly reduced compared to the classical model. Second, we apply the obtained drag force to calculate the terminal velocities of interstellar grains by strong radiation sources such as supernovae and active galactic nuclei (AGNs). We find that grains can be accelerated to relativistic velocities by very luminous AGNs. We then quantify the deceleration of relativistic spacecraft proposed by the Breakthrough Starshot initiative due to gas drag on a relativistic lightsail. We find that the spacecraft's slowing down is negligible because of the suppression of gas drag at relativistic velocities, suggesting that the lightsail may be open for communication during its journey to $\alpha$ Centauri without causing a considerable delay. Finally, we show that the damage to relativistic thin lightsails by interstellar dust is a minor effect.
\end{abstract}

\section{Introduction\label{sec:intro}}
The motion of dust grains through the ambient gas inevitably experiences drag force, which acts in the opposite direction to the direction of grain motion and leads to the grain slowing down. 

Radiation pressure is a main acceleration mechanism for dust grains, leading to the motion of grains through the ambient gas. Radiation pressure from late-type stars (e.g., Asymptotic Giant Branch (AGB), post-AGBs, and planetary nebulae (PNe)) is found to produce grain motion in circumstellar envelopes (\citealt{1976ApJ...205..144G}; \citealt{1993ApJ...410..701N}).  In addition, fast modes of magnetohydrodynamic turbulence are found to accelerate charged dust grains to supersonic velocities (\citealt{2003ApJ...592L..33Y}; \citealt{Yan:2004ko}; \citealt{Hoang:2012cx}). Moreover, transit time damping is found to be an efficient acceleration mechanism \citep{Hoang:2012cx}. The aforementioned astrophysical conditions can produce grain motion at non-relativistic velocities ($v\ll 0.1c$). 

\cite{Spitzer:1949bv} first noticed that radiation pressure from very strong radiation sources, such as supernovae (SNe), can accelerate grains to speeds close to that of light. The ejection of dust grains by SNe and galactic winds is believed to be the main process for the enrichment of metal in intergalactic medium \citep{2005MNRAS.358..379B}. Diffusive shocks from supernova remnants (SNRs) are also found to be important for acceleration of charged grains to very high velocities (\citealt{1997ApJ...487..197E}; \citealt{2009ApJ...701.1865G}). All in all, the motions of grains in astrophysical conditions can occur at a wide range of velocities, from thermal to relativistic motions. 

The classical model of gas drag (\citealt{Epstein:1924tc}; \citealt{Baines:1965p3201}; \citealt{1979ApJ...231...77D}) is widely used in astrophysics. However, the original model becomes inapplicable for grains moving at very high velocities. Indeed, the classical model assumes that, upon collisions, impinging atoms transfer their entire momentum to the grain via sticky collisions or even deposit twice its momentum to the grain via secular reflection. Such an assumption becomes invalid when the speed of impinging atoms is sufficiently large such that their penetration depth is much larger than the grain diameter. The goal of this paper is first to quantify the gas drag force for grains moving at relativistic velocities, so-called {\it relativistic gas drag}.\footnote{The diameter of dust grains in the diffuse interstellar medium is much smaller than the mean free path. Thus, gas drag is well within the Epstein regime, whereas the Stoke regime occurs for grains larger than a few cm.} {Note that a related problem of relativistic radiation drag on a planar surface moving through an isotropic radiation field is studied in \cite{Lee:2016jb}.}

A new model for relativistic gas drag has important implications for finding the terminal velocities of grains accelerated by very strong radiation fields, which has important implication for dust destruction in the ISM and enrichment of intergalactic medium \citep{2005MNRAS.358..379B}. Indeed, the terminal velocities achieved by radiation pressure is obtained by the balance between the radiation force and drag force. The classical drag increases rapidly with the grain velocity as $v^{2}$ (see \citealt{2011ApJ...732..100D}), which can reduce significantly the grain terminal velocities. 

Interestingly, the Breakthrough Starshot initiative\footnote{https://breakthroughinitiatives.org/Initiative/3} aims to launch gram-scale spacecraft with miniaturized electronic components (such as camera, navigation, and communication systems) to relativistic velocities ($v\sim 0.2c)$. This will enable the spacecraft to reach the nearest stars, like $\alpha$ Centauri (a distance of 1.34 pc), within a human lifetime. Powerful laser beams will be used to propel a thin, highly reflective lightsail attached to the spacecraft to relativistic speeds. Previous studies on the acceleration of relativistic lightsail by laser beams ignore the effect of gas drag (\citealt{2016SPIE.9981E..06K}; \citealt{2017AJ....153..277K}). Moreover, the motion of the relativistic lightsail in the interstellar medium (ISM) will be subject to relativistic gas drag, such that the spacecraft will be decelerated. This issue will be quantified in this paper.

In this paper, we will first quantify the gas drag on dust grains moving at relativistic velocities in Section \ref{sec:drag}. In Section \ref{sec:result} we will employ the obtained relativistic drag force to calculate the velocities of grains accelerated by luminous astrophysical sources, such as SNe and AGNs. We will estimate the deceleration of the relativistic spacecraft with an open lightsail and damage to the lightsail by interstellar dust in Section \ref{sec:starshot}. Our main conclusions are presented in Section \ref{sec:discuss}. 

\section{Gas Drag: Low and high energy regimes}\label{sec:drag}

\subsection{Low energy regime: classical drag}
Let us consider an atomic gas of density $n$ and mass $m$ for simplicity. A neutral grain of size $a$ moving with low velocity $v$ experiences a drag force, given by the classical formula:
\begin{eqnarray}
F_{\rm drag} = 2\pi a^{2}kT\left(n G_{0}(s)\right),\label{eq:Fdrag_low}
\end{eqnarray}
where $T$ is the gas temperature, and 
\begin{eqnarray}
s = \left(\frac{mv^{2}}{2kT}\right)^{1/2}\equiv v/v_{T},\\
G_{0}(s) \approx \frac{8s}{3\sqrt{\pi}}\left(1 + \frac{9\pi}{64}s^{2}\right)^{1/2},\end{eqnarray}
with $v_{T}$ being the thermal velocity (see \citealt{2011ApJ...732..100D}).

At high velocities, (i.e., $s\gg 1$), Equation (\ref{eq:Fdrag_low}) becomes
\begin{eqnarray}
F_{\rm drag} =n\pi a^{2}(mv^{2}),\label{eq:dragcoll}
\end{eqnarray}
which has a linear scaling with the kinetic energy $E$.

\subsection{High Energy Regime}\label{eq:highE}
For rapidly moving grains, impinging atoms can pass through the grain, transferring only part of their momentum.\footnote{For energetic atom, the electron and nucleon interact independently with the target. Here we disregard the electron due to its much smaller momentum compared to nucleon.}  

Indeed, the penetration depth of impinging protons is approximately equal to $R_{\rm H}(E)=1 (3\g\cm^{-3}/\rho)(E/100~\rm keV) \mum$ with $\rho$ being the mass density of dust (\citealt{1979ApJ...231...77D}). It indicates that for micron-sized grains (or thin lightsails), the penetration depth is much larger than the grain diameter at $E> 100$ keV, such that the classical formula is not applicable. In this section, we will drive a new drag formula for this regime.

\subsubsection{Non-relativistic case}
First, we consider the non-relativistic regime, i.e., $v<0.1c$. To derive the drag force for the high energy regime, we first need to find the momentum (or kinetic energy) that the impinging ion transfers to the grain during its passage. The kinetic energy loss of an ion after passing through the grain is given by
\begin{eqnarray}
\Delta E = \frac{4a}{3}\frac{dE}{dx},\label{eq:deltaE}
\end{eqnarray}
where $dE/dx$ is the stopping power, and we have approximated the spherical grain as a slab of thickness $4a/3$.

Let $E, p$ be the kinetic energy and momentum of the incident ion. The kinetic energy and momentum of the ion exiting the grain, denoted by $E'$ and $p'$, are governed by the law of energy conservation:
\begin{eqnarray}
E' = E - \Delta E,\\
p'^{2} = p^{2} - 2m \Delta E.
\end{eqnarray}

It is straightforward to derive the decrease in the ion momentum due to the passage through the grain:
\begin{eqnarray}
\Delta p &=& p - p' = \frac{2m\Delta E}{p + p'}=\frac{2m\Delta E}{p+ (p^{2} - 2m\Delta E)^{1/2}}.
\end{eqnarray}

For interstellar grains with $a<1\mum$ and energetic ions, we have $\Delta E \ll p^{2}/2m$. Thus, the above equation can be rewritten as
\begin{eqnarray}
\Delta p =  \frac{2mp\Delta E}{2p^{2} - m\Delta E} \approx  \frac{\Delta E}{v}.\label{eq:dp}
\end{eqnarray}
 
With the rate of collisions $R_{\rm coll}=nv\pi a^{2}$, the total momentum that gas atoms transfer to the grain per second, namely drag force, is
\begin{eqnarray}
F_{\rm drag}= R_{\rm coll} \Delta p = n \pi a^{2} \Delta E.\label{eq:Fdrag1}
\end{eqnarray}

The stopping power of proton impact can be approximated as
\begin{eqnarray}
\frac{dE}{dx} = \frac{2S_{m}(E/E_{m})^{\eta}}{1+(E/E_{m})},\label{eq:dEdx_aprx}
\end{eqnarray}
where $\eta$ is the slope, $E_{m}=100$ keV and $S_{m}$ is the stopping power at $E=E_{m}$. For graphite, we find that $\eta=0.2$ and $S_{m}= 1.8\times 10^{6}$ keV/cm. For quartz material, $\eta=0.25$ and $S_{m}=1.3\times 10^{6}$ keV/cm.

\begin{figure}
\includegraphics[width=0.5\textwidth]{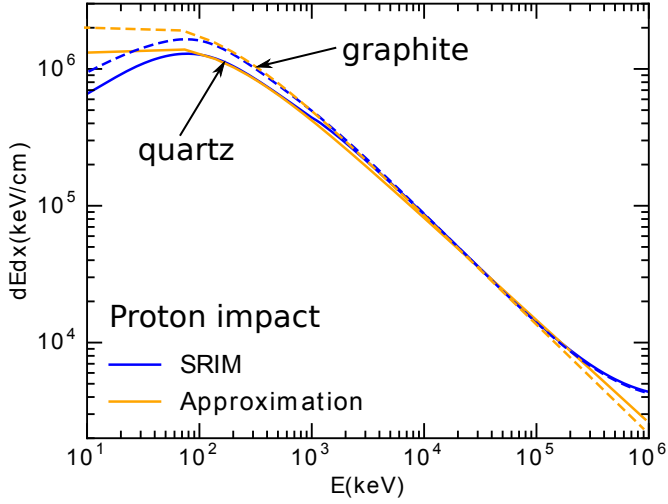}
\caption{Stopping power as a function of the proton kinetic energy for quartz (solid lines) and graphite (dashed lines), obtained from numerical calculations with the SRIM code and our analytical approximation. An excellent fit is achieved for $E\sim 10^{2}-10^{5}$ keV.}
\label{fig:dEdx}
\end{figure}

Figure \ref{fig:dEdx} shows an excellent fit of the analytical formula to the exact results calculated with the SRIM code for both graphite and quartz materials (\citealt{2010NIMPB.268.1818Z}) for $E\sim 10^{2} - 10^{5} {\rm keV}$.

From Equations (\ref{eq:deltaE}), (\ref{eq:Fdrag1}), and (\ref{eq:dEdx_aprx}), we obtain an analytical formula for the drag force:
\begin{eqnarray}
F_{\rm drag}&=&n\pi a^{2}(4a/3)\left(\frac{2S_{m}(E/E_{m})^{\eta}}{1+(E/E_{m})}\right),\label{eq:Fdrag_nrel}
\end{eqnarray}
which scales as $1/E^{1-\eta}$ with $\eta\sim 0.2$ (or 0.25 for quartz material) for $E\gg E_{m}$. {This new drag force has a different scaling from the classical drag with $F_{\rm drag}\propto E$ (see Eq. \ref{eq:dragcoll}), and is several orders of magnitude lower than predicted by the classical model at the same energy $E>E_{m}$.}

\subsubsection{Relativistic case}
Now, let consider the case where dust grains are moving through the gas at relativistic velocities, i.e. $v>0.1c$ or $\beta=v/c>0.1$. The Lorentz factor is $\gamma=1/(1-\beta^{2})^{1/2}$.

The relativistic momentum of an incident particle of rest mass $m$ is 
\begin{eqnarray}
p = mc\gamma \beta,
\end{eqnarray}
and its kinetic energy $K=mc^{2}(\gamma -1)$.

The gas density in the reference frame fixed to the grain is $\gamma n$ (see \citealt{2015ApJ...806..255H}, hereafter HLS15). Then, the force acting on the grain is given by
\begin{eqnarray}
F_{\rm drag}= \gamma n\beta c \pi a^{2} \Delta p.
\end{eqnarray}

Following the same procedure as in the non-relativistic case, we obtain the kinetic energy and Lorentz factor of the ion after the collision given by:
\begin{eqnarray}
K- K' = mc^{2}(\gamma -\gamma')=\Delta E,
\end{eqnarray} 
which yields
\begin{equation}
\gamma'= \gamma -\frac{\Delta E}{mc^{2}}.\label{eq:gammaf}
\end{equation}

The decrease in the ion momentum is calculated by
\begin{eqnarray}
\Delta p = mc\left(\gamma\beta - \gamma'\beta'\right),\label{eq:dp_rel}
\end{eqnarray}
which is valid for arbitrary value of $\beta$.

Thus, the relativistic drag force is equal to
\begin{eqnarray}
F_{\rm drag}= nmc^{2}\pi a^{2}\gamma^{2}\beta^{2}
\left(1 - \frac{\gamma'\beta'}{\gamma\beta}\right).\label{eq:Fdrag_rel}
\end{eqnarray}

\begin{figure}
\includegraphics[width=0.5\textwidth]{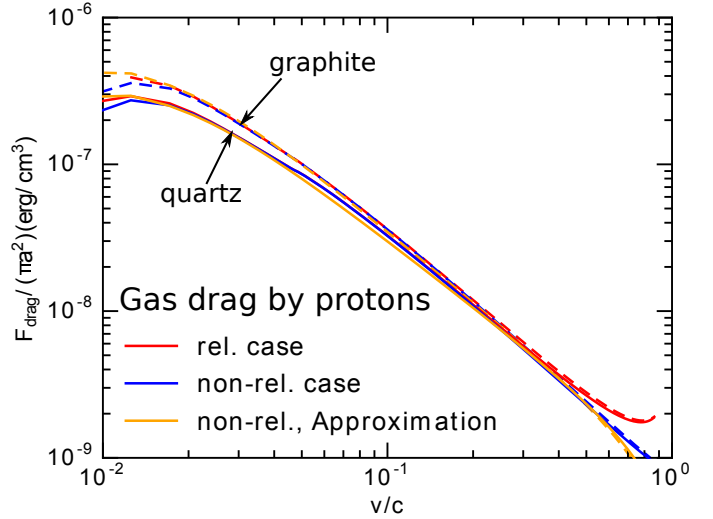}
\caption{Drag force as a function of the grain velocity computed for the relativistic (red line) and non-relativistic (blue line) cases. The drag force obtained with our analytical formula is also shown for comparison (orange line). The drag force rapidly decreases for $v>0.01c$. Quartz (solid lines) and graphite (dashed lines) grains of $a=0.1\mum$ are considered.}
\label{fig:Fdrag}
\end{figure}

Figure \ref{fig:Fdrag} shows the drag forces calculated using the non-relativistic (Equation \ref{eq:Fdrag1}, blue line), analytical formula (Equation \ref{eq:Fdrag_nrel}, orange line), and relativistic (Equation \ref{eq:Fdrag_rel}, red line) formulae with the stopping power $dE/dx$ computed with the SRIM code from \cite{2017ApJ...837....5H}. A good agreement in the drag force from the different approaches is observed for $v<0.5c$. 

{The problem of gas drag on relativistic lightsails was analyzed in \cite{1990JSpRo..27...48M}. However, the authors adopted an approximation for $\Delta p$ (their Equation 37), which gives a different form of gas drag that decreases with $v$ much slower than our formula. Note that a charged grain also experiences Coulomb drag due to electrostatic interaction with the plasma. Because the cross-section of Coulomb interaction scales as $1/v^{2}$, the Coulomb drag appears to be negligible for relativistic grains. Finally, we have treated a spherical grain as a slab. In reality, a spherical grain moving at relativistic velocities would experience Lorentz contraction, but this effect is small for $\gamma<2$ and thus ignored in this paper.}

\section{Grain Acceleration by strong radiation field and Numerical results}\label{sec:result}
In this section, we apply our new drag force to calculate terminal velocities of grains accelerated by strong radiation sources.

\subsection{Radiation Pressure Force}

Supernovae (type Ia and type II) have typical luminosity $L_{\rm bol}\sim 10^{8}L_{\odot}$, for which grains are unlikely accelerated to $v\gg 0.1c$. In this {\it non-relativistic} regime, the radiation pressure force acting on a stationary grain at distance $r$ from the central source is given by the usual formula:
\begin{eqnarray}
F_{\rm rad}=\int d\nu c\frac{u({\nu})}{h\nu} \frac{h\nu}{c} Q_{\rm pr,\nu}\pi a^{2} =
u_{\rm rad}\langle Q_{\rm pr}\rangle \pi a^{2},\label{eq:Frad}
\end{eqnarray}
where $u({\nu})=L_{\nu}/(4\pi r^{2}c)$, $u_{\rm rad}=L_{\rm bol}/(4\pi r^{2}c)$ with  $L_{\nu}$ being the specific luminosity, $Q_{\rm pr,\nu}$ is the radiation pressure efficiency by photon of frequency $\nu$, and 
\begin{eqnarray}
\langle Q_{\rm pr}\rangle = \frac{\int Q_{\rm pr, \nu} u(\nu)d\nu}{\int u(\nu)d\nu},\label{eq:Qpr_avg}
\end{eqnarray}
is the radiation pressure efficiency averaged over the radiation spectrum seen by the grain (see HLS15 for more details). 

Quasars or Seyfert galaxies have a typical luminosity $L_{\rm bol} \sim 10^{13} L_{\odot}$, which are expected to accelerate grains to $v\gg 0.1c$ (see HLS15). For this {\it relativistic} regime, the Doppler effect must be taken into account. As a result, the radiation pressure force is given by 
\begin{eqnarray}
F_{\rm rad} = \int d\nu' c\frac{u({\nu'})}{h\nu'}\frac{h\nu'}{c} Q_{\rm pr,\nu'}\pi a^{2} =u'_{\rm rad}\langle Q_{\rm pr}\rangle_{\gamma}\pi a^{2},~~~\label{eq:Fradrel}
\end{eqnarray}
where $\nu'$ is the frequency of photon and $u'_{\rm rad}=L'_{\rm bol}/(4\pi r^{2}c)$ is the energy density, and the prime denotes the physical quantity in the grain reference frame. Here $\langle Q_{\rm pr}\rangle_{\gamma}$ is given by Equation (\ref{eq:Qpr_avg}) with $\nu$ replaced by $\nu'$, which results in its dependence on $\gamma$ (see HLS15).

\subsection{Equation of Motion}

For stellar radiation sources in which grains are unlikely accelerated to $v>0.5c$, the equation of {\it non-relativistic} motion can be described by
\begin{eqnarray}
\frac{m_{\rm gr}dv}{dt} &=& F_{\rm rad} - F_{\rm drag},\label{eq:dvdt}\\
\frac{dr}{dt} &=& v,\label{eq:drdt}
\end{eqnarray}
where $m_{\rm gr}$ is the grain mass, $F_{\rm drag}$ and $F_{\rm rad}$ are given by Equations (\ref{eq:Fdrag_nrel}) and (\ref{eq:Frad}), respectively. 

For AGNs or Seyfert galaxies, it is expected that grains can be accelerated to $v>0.5c$ due to high radiation intensity. Therefore, the equation of {\it relativistic} motion for the radial velocity component of {\it relativistic particles} takes the following form:
\begin{eqnarray}
\frac{dp'_{\rm gr}}{dt'}&\equiv&\frac{m_{\rm gr}cd\tilde{u}}{dt'}=F_{\rm rad}w\left(1-w\tilde{u}\right) - F_{\rm drag},\label{eq:duds}\\
\frac{dr}{dt'}&=&c\tilde{u},
\end{eqnarray}
where $p'_{\rm gr}$ is the grain relativistic momentum, $dt'$ is the proper time measured in the grain frame, $\tilde{u}=\gamma \beta$, and $w=\gamma-\tilde{u}=\gamma(1-\beta)$ accounts for the redshift by Doppler effect (\citealt{1937MNRAS..97..423R}; \citealt{1971Ap&SS..13...70N}; HLS15). Here, $F_{\rm drag}$ and $F_{\rm rad}$ are given by Equations (\ref{eq:Fdrag_rel}) and (\ref{eq:Fradrel}), respectively. 

Substituting $dr= c\beta dt = c\tilde{u} dt'$ into Equation (\ref{eq:duds}), we obtain
\begin{eqnarray}
\frac{d\tilde{u}}{dr}=\left[\frac{L'_{\rm bol}}{4\pi r^{2}c}\langle Q_{\rm pr}\rangle_{\gamma}\pi a^{2}f(\tilde{u})-F_{\rm drag}\right]
\left(\frac{1}{m_{\rm gr}c^{2}\tilde{u}}\right),\label{eq:dudr}
\end{eqnarray}
where $f(\tilde{u})=\left(\gamma-2 \tilde{u} -2\tilde{u}^{3} + 2\gamma \tilde{u}^{2}\right)$.

Above, we have ignored the subdominant contribution of gravitational force and Lorentz force during the acceleration stage.

\subsection{Grain acceleration by SNe}
We find the grain terminal velocities due to SNe radiation by solving the equation of motion for the non-relativistic case (Equation \ref{eq:dvdt}). We consider grains in a cloud of $n=10^{3}\cm^{-3}$ and $T=20 \K$, and subject to SNe radiation of total luminosity $L_{\rm bol}\sim 10^{6}-10^{9}L_{\odot}$. Note that a SN Ia has the luminosity declining from its peak of $\sim 10^{9}L_{\odot}$ to $10^{7}L_{\odot}$ after $\sim$ 300 days. Here, we assume the luminosity is constant and consider several values for it.

Initially, grains are assumed to be located at the sublimation distance, $r_{\rm sub}$, from the central source, which is given by
\begin{eqnarray}
r_{\rm sub}=\left(\frac{L_{\rm UV}}{5\times 10^{12}L_{\odot}}\right)^{1/2}\left(\frac{T_{\rm sub}}{180
0\K}\right)^{-5.6/2} {\rm pc},\label{eq:rsub}
\end{eqnarray}
where  $L_{\rm UV}$ is the luminosity in the optical and UV, which is roughly one half of the bolometric luminosity, and $T_{\rm sub}$ is the dust sublimation temperature (see \citealt{1995ApJ...451..510S}). In the following, we adopt $T_{\rm sub}=1500\K$ and $1800$K for silicate and graphite grains (see \citealt{1989ApJ...345..230G}).

\begin{figure}
\includegraphics[width=0.5\textwidth]{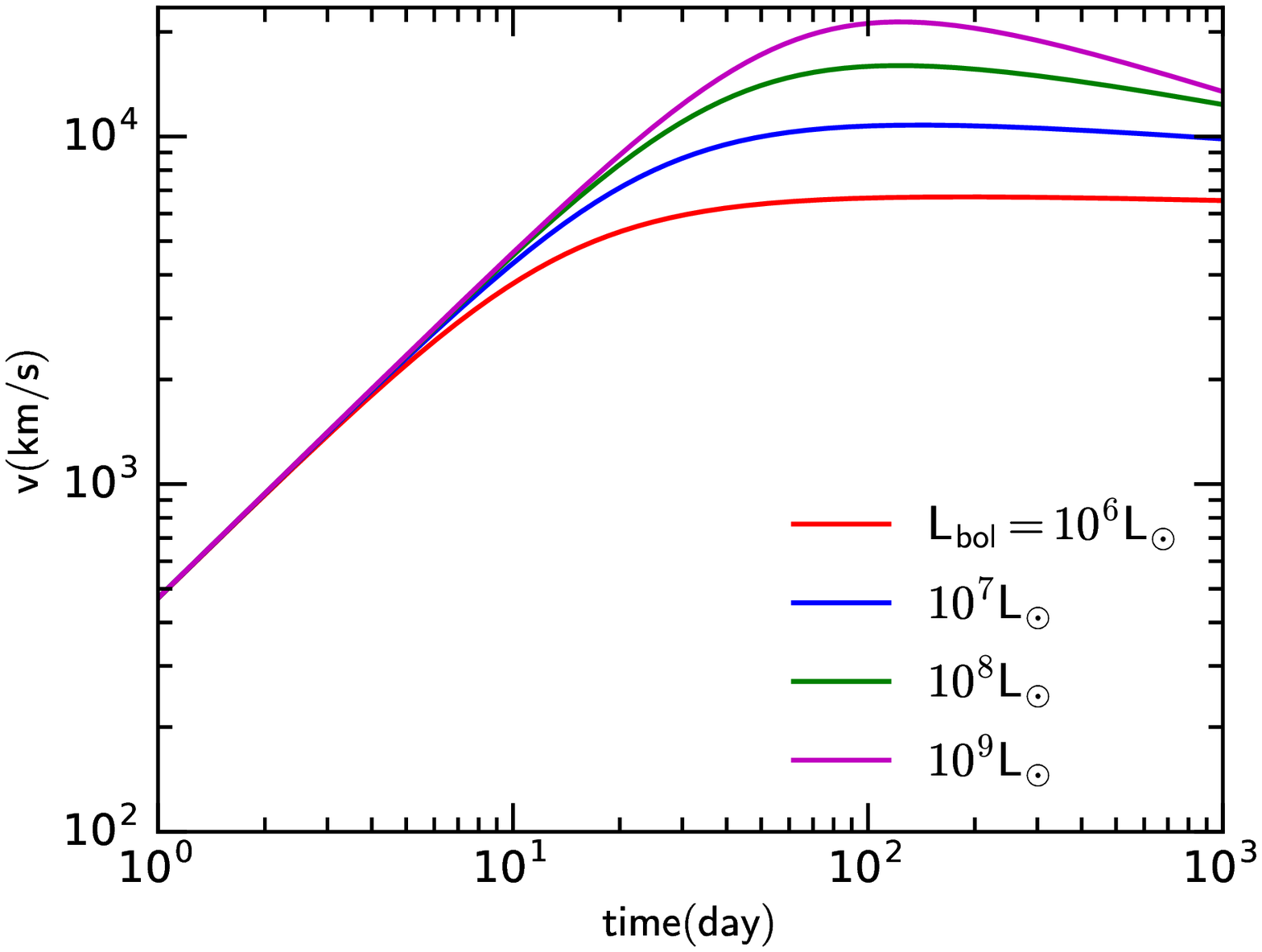}
\includegraphics[width=0.5\textwidth]{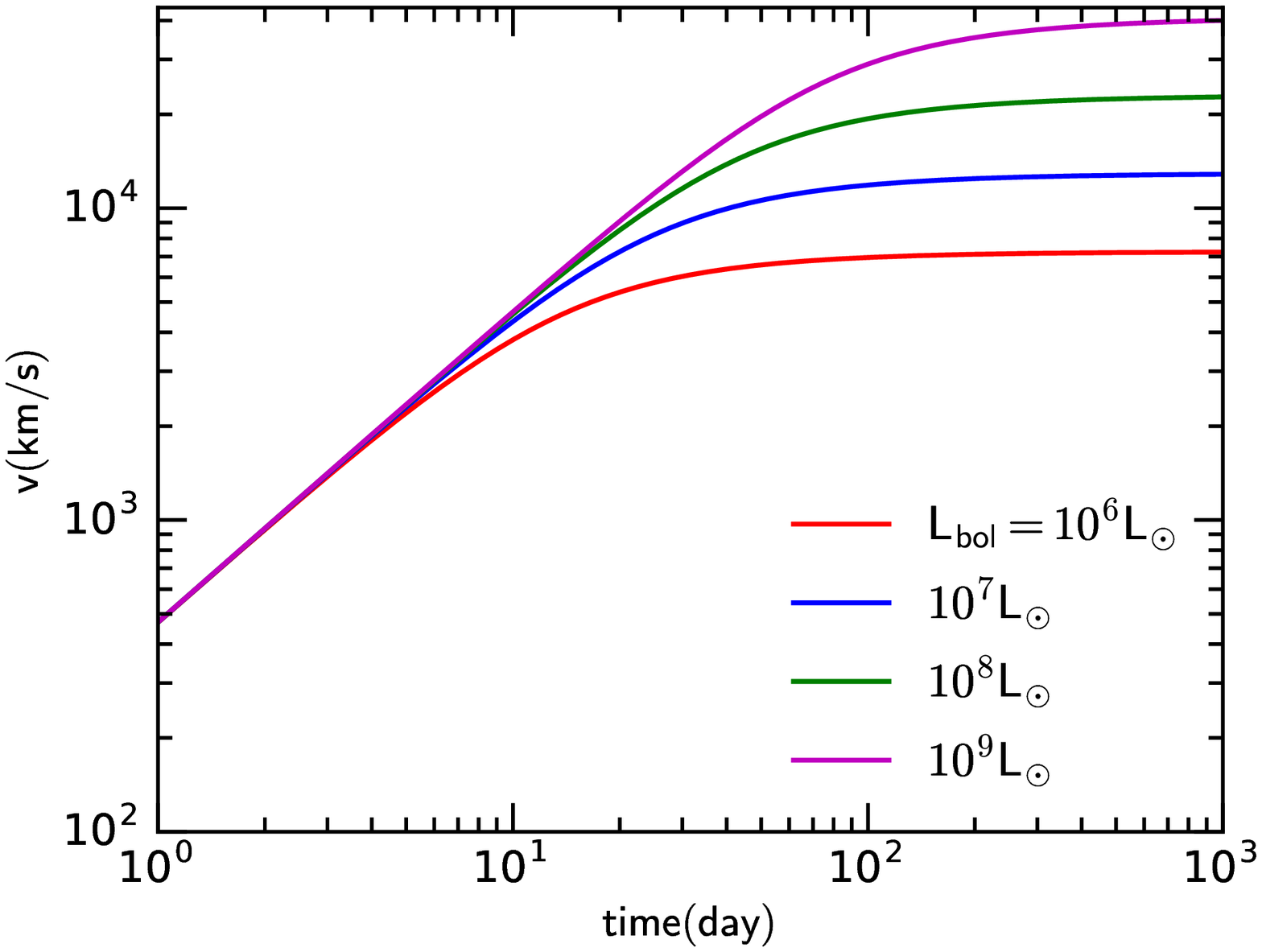}
\caption{Grain velocities accelerated by SNe as a function of time obtained with the classical drag force ({\bf upper panel}) and new drag force ({\bf lower panel}) for the different luminosity. The classical drag force induces the decrease of $v$ after the peak (upper panel). Graphite grains with size $a=0.1\mum$ are considered.}
\label{fig:Vrad_SN}
\end{figure}

The upper panel of Figure \ref{fig:Vrad_SN} shows the grain velocities as a function of time using the classical drag force. The grain velocities increase rapidly to their maximum values of $v_{\rm max}>10^{4}$ km/s for $L_{\rm bol}>10^{6}L_{\odot}$. Beyond the peaks, the grain velocities decline due to the rapid increase of the gas drag as $v^{2}$. In the lower panel where the new drag force is used, the grain velocities continue to rise and achieve its terminal values. The maximum grain velocity induced by SNe radiation is $v\sim 0.13c$ for $L_{\rm bol}=10^{9}L_{\odot}$.

\subsection{Grain acceleration by AGNs}
\begin{figure}
\includegraphics[width=0.5\textwidth]{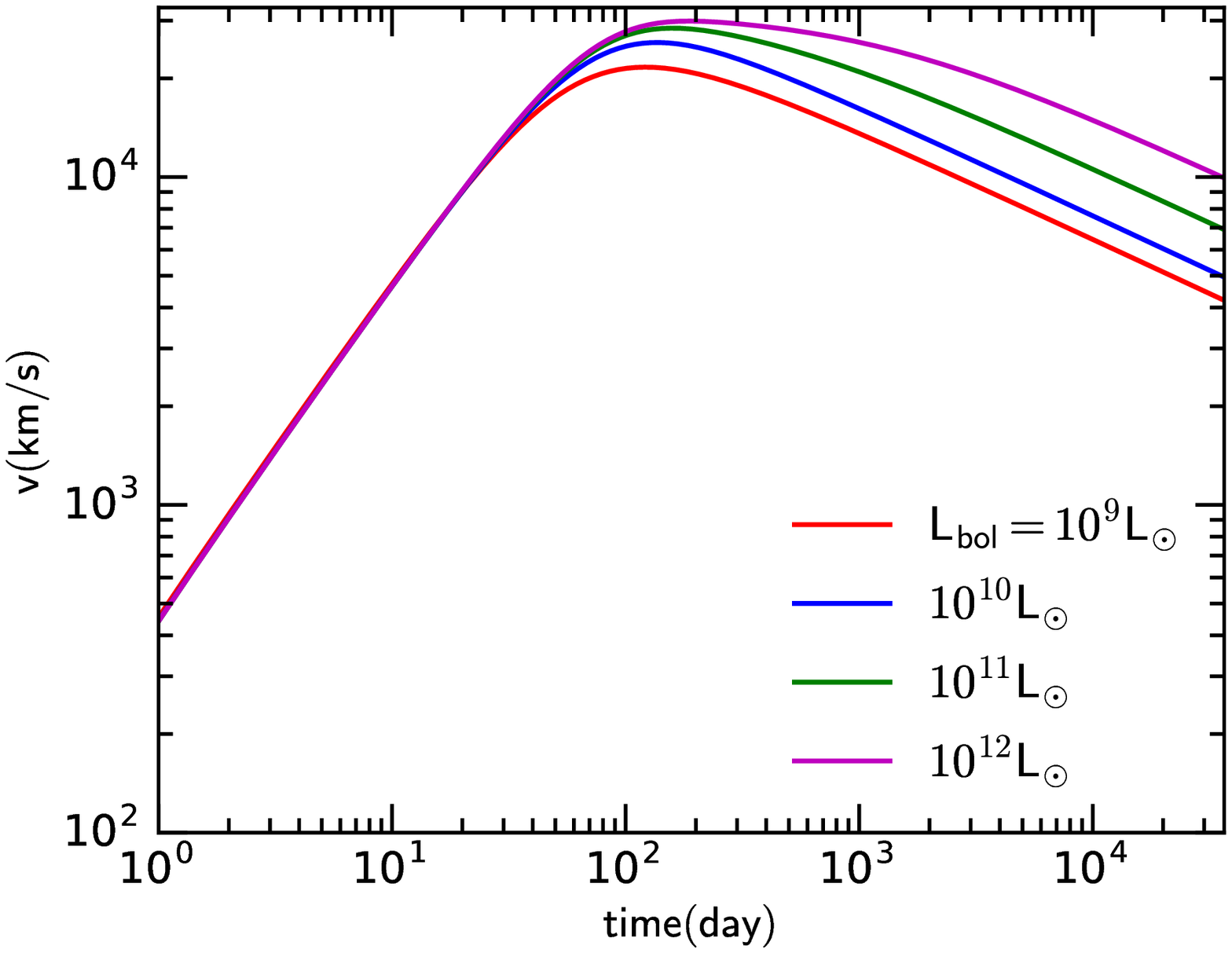}
\includegraphics[width=0.5\textwidth]{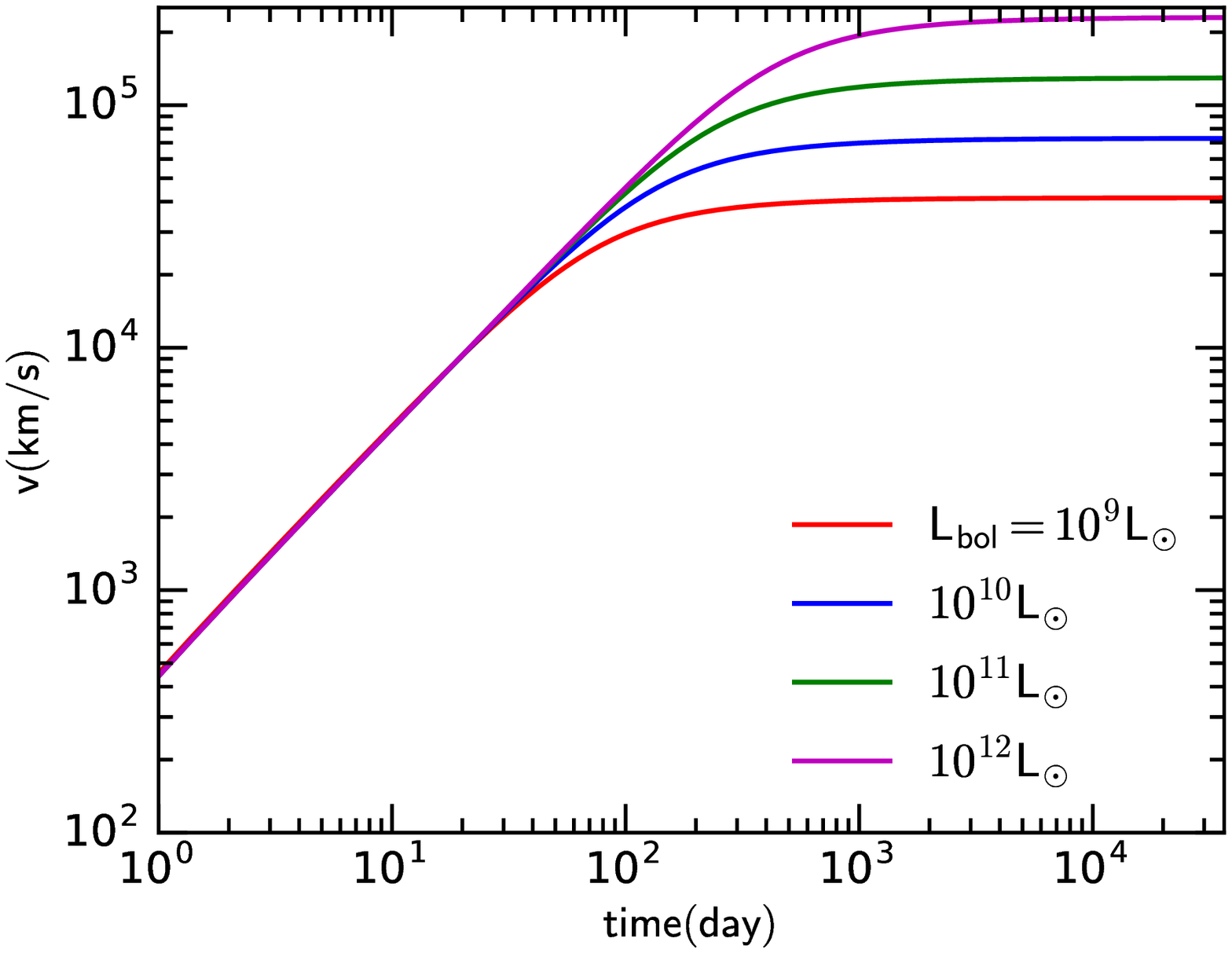}
\caption{Same as Figure \ref{fig:Vrad_SN}, but for AGNs with $L_{\rm bol}=10^{9}-10^{12}L_{\odot}$. Grains can be accelerated to relativistic velocities of $v>0.1c$ for $L_{\rm bol}=10^{12}L_{\odot}$.}
\label{fig:Vrad_AGN}
\end{figure}

We first consider AGNs with typical luminosity of $L_{\rm bol}=10^{9}-10^{12}L_{\odot}$, for which $v>0.5c$ is not expected. Thus, we find the grain velocities by solving the non-relativistic equations of motion. The resulting velocities are shown in Figure \ref{fig:Vrad_AGN}. We see a similar trend as in the SNe case, but the terminal velocities are much higher due to larger $L_{\rm bol}$, as expected.

The most luminous quasar discovered to date has $L_{\rm bol}\sim 4.3\times 10^{14}L_{\odot}$ (\citealt{Wu:kr}), which is expected to accelerate grains to $v> 0.5c$. Thus, we can employ the relativistic equations of motion (Equation \ref{eq:dudr}) to find grain terminal velocities for $L_{\rm bol}$ upto $10^{15}L_{\odot}$. First, we substitute $L'_{\rm bol}=L_{\rm bol}\gamma^{2}(1-\beta)^{2}$ (see HLS15). We also compute $\langle Q_{\rm pr}\rangle_{\gamma}$ for different $\gamma$ using a typical radiation spectrum of unobscured AGNs, which will be used to interpolate for $\langle Q_{\rm pr}\rangle_{\gamma}$ and calculations of $F_{\rm rad}$. Then, we solve Equation (\ref{eq:dudr}) numerically for $\tilde{u}$ as a function of $r$ with initial radius $r_{i}=r_{\rm sub}$ and final radius $r_{f}=20r_{\rm sub}$.

\begin{figure}
\includegraphics[width=0.5\textwidth]{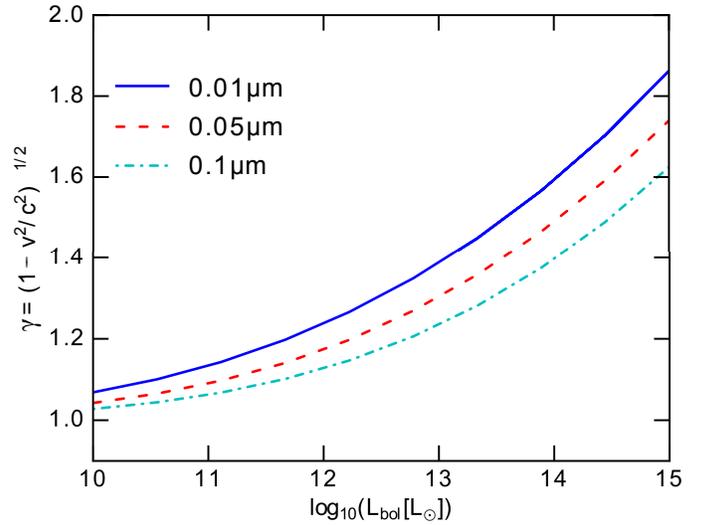}
\caption{Lorentz factor $\gamma=\left(1-v^{2}/c^{2}\right)^{-1/2}$ as a function of the luminosity for three different grain sizes. Relativistic grains with $\gamma\sim 1.8$ ($v\sim 0.85c$) are achieved by the most luminous AGN.}
\label{fig:gamrad}
\end{figure}

The terminal velocities represented through $\gamma$ for the different grain sizes are shown in Figure \ref{fig:gamrad}. It can be seen that, for brightest quasars or Seyfert galaxies of $L_{\rm bol}\sim 10^{14}-10^{15}L_{\odot}$, radiation pressure can accelerate dust grains to relativistic velocities with $v\sim 0.75c-0.85c$ ($\gamma\sim 1.5-1.8$), in agreement with the results from HLS15 where the drag force is disregarded. 

\section{Implications for relativistic lightsails}\label{sec:starshot}
\subsection{Deceleration of relativistic lightsails}
The Breakthrough Starshot initiative plans to use powerful laser beams to propel a spacecraft attached to a thin, highly reflective lightsail to $v\sim 0.2c$. The lightsail is recommended to be folded to minimize the effect of gas drag during the flight. Nevertheless, whether interstellar gas can lead to a significant deceleration of the spacecraft is still unknown. 

Using Equation (\ref{eq:Fdrag_nrel}), we estimate the slowing down of a lightsail of area $A_{\rm sail}$ and thickness $l$ as follows:
\begin{eqnarray}
\frac{\Delta v}{v} &=& \frac{F_{\rm drag}\Delta t}{Mv}=\frac{nA_{\rm sail}\Delta E\Delta t}{Mv}\simeq\left(\frac{2.5\times 10^{-6}}{M}\right)\nonumber\\
&\times&\left(\frac{N_{\rm H}}{10^{18}\cm^{-2}}\right)\left(\frac{0.2c}{v}\right)^{2.8}\left(\frac{A_{\rm sail}}{1\rm m^{2}}\right)\left(\frac{l}{1\mum}\right),~~~
\end{eqnarray}
where $M$ is the total mass of the spacecraft and the lightsail, $N_{\rm H}$ is the gas column swept by the lightsail, and $E=m_{\rm H}v^{2}/2$.

For a lightsail of $A_{\rm sail}=1 {\rm m}^{2}$ and $l=1\mum$, and $M= 2$g, we get $\Delta v/v=10^{-6}$ for $N_{\rm H}=10^{18}\cm^{-2}$, i.e., by the time the spacecraft reaches $\alpha$ Centauri \citep{1996ApJ...463..254L}. Thinner lightsails of 10 nm experience much weaker deceleration. Therefore, the slowing down of the spacecraft by interstellar gas leads to a delay in the spacecraft arrival by a couple of minutes, which is negligible. Note that the lightsail inevitably gets charged by collisions with interstellar gas, which can lead to the deflection of the spacecraft from the intended target and spacecraft oscillation \citep{Hoang:2017um}.

Finally, we note that previous studies on the acceleration of relativistic lightsail by laser beams ignore the effect of gas drag (\citealt{2016SPIE.9981E..06K}; \citealt{2017AJ....153..277K}). In the light of our new result, the effect of gas drag will be negligible because the spacecraft would be rapidly accelerated to very high velocity at which the drag force is substantially reduced.

\subsection{Damage of relativistic lightsail by interstellar matter}
{Thin lightsails will be damaged by interstellar gas and dust. At relativistic speeds, the damage by gas is not important because atoms pass through the thin sail. The issue of damage of very thin lightsails by dust impact was studied in \cite{2000JSpRo..37..526E}, where the authors show that each dust grain can produce a hole of diameter not much larger than the grain diameter at $v>0.1$c.

Let us estimate the total damage caused by interstellar dust towards $\alpha$ Centauri. We assume that a dust grain will produce a hole of the diameter of twice that of the grain, which is an upper limit for lightsails moving at $v>0.1c$ (see \citealt{2000JSpRo..37..526E}). Then, the fraction of the lightsail surface area damaged after it swept a gas column of $N_{\rm H}$ is given by
\begin{eqnarray}
f_{\rm S,sail} &=& \int_{a_{\rm min}}^{a_{\rm max}} \pi(2a)^{2} A_{\rm MRN}a^{-3.5} da N_{\rm H}\nonumber\\
&\simeq & 8\pi a_{\rm min}^{-0.5}A_{\rm MRN}N_{\rm H}\nonumber\\
&\sim& 0.01\left(\frac{a_{\rm min}}{3.5\AA}\right)^{-1/2}\left(\frac{N_{\rm H}}{10^{18}\cm^{-2}}\right),
\end{eqnarray}
where we have assumed that the size distribution of interstellar dust follows a power law distribution from \cite{Mathis:1977p3072} with $a_{\rm min}=3.5$\AA, $a_{\rm max}=0.25\mum$, and $A_{\rm MRN}=10^{-25.16}\cm^{-2.5}$. Thus, upto $1\%$ of the lightsail will be damaged by dust impact by the time the spacecraft reaches $\alpha$ Centauri given by $N_{\rm H}\sim 10^{18}\cm^{2}$.}

\section{Conclusions}\label{sec:discuss}

We have derived a new formula of drag force experienced by grains moving at high velocities through the interstellar gas, for both non-relativistic ($v< 0.1c$) and relativistic ($v\gg 0.1c$) regimes. The relativistic drag force is found decreasing with the kinetic energy of impinging ions as $1/E^{0.8}$ for $E\gg E_{m}=100$ keV, which is different from the linear scaling of the classical drag formula.
 
Using the new drag formula, we calculated the terminal velocities of grains accelerated by radiation pressure from SNe and AGNs. We found that grains can be accelerated to relativistic velocities by very strong radiation sources, such as Seyfert galaxies and Quasars. Such relativistic dust grains were once referred to explain ultrahigh energy cosmic rays, but it is shown in \citep{2015ApJ...806..255H} that relativistic grains would be destroyed rapidly in the ISM by Coulomb explosions.

Finally, we showed that the slowing down of relativistic lightsails of the type envisioned by Breakthrough Starshot due to interstellar gas drag is negligible thanks to the suppression of gas drag at relativistic velocities. Thus, the lightsail can be open for communication and navigation during its journey to $\alpha$ Centauri without resulting in a considerable delay. We also evaluated the damage of thin relativistic lightsails by interstellar dust.

\acknowledgments
{We thank an anonymous referee for useful comments and suggestions that improve our paper.} The original idea for relativistic gas drag was born during our stay at the Ruhr-Universit$\ddot{\rm a}$t Bochum as an Alexander von Humboldt Fellow, but the completion of this work is inspired by the Breakthrough Starshot initiative. This work was supported by the Basic Science Research Program through the National Research Foundation of Korea (NRF), funded by the Ministry of Education (2017R1D1A1B03035359).


\bibliography{ms.bbl}

\begin{thebibliography}{29}
\expandafter\ifx\csname natexlab\endcsname\relax\def\natexlab#1{#1}\fi

\bibitem[{Baines {et~al.}(1965)Baines, Williams, \&
  Asebiomo}]{Baines:1965p3201}
Baines, M.~J., Williams, I.~P., \& Asebiomo, A.~S. 1965, \mnras, 130, 63

\bibitem[{Bianchi \& Ferrara(2005)}]{2005MNRAS.358..379B}
Bianchi, S., \& Ferrara, A. 2005, \mnras, 358, 379

\bibitem[{Draine(2011)}]{2011ApJ...732..100D}
Draine, B.~T. 2011, \apj, 732, 100

\bibitem[{Draine \& Salpeter(1979)}]{1979ApJ...231...77D}
Draine, B.~T., \& Salpeter, E.~E. 1979, \apj, 231, 77

\bibitem[{Early \& London(2000)}]{2000JSpRo..37..526E}
Early, J.~T., \& London, R.~A. 2000, Journal of Spacecraft and Rockets, 37, 526

\bibitem[{Ellison {et~al.}(1997)Ellison, Drury, \& Meyer}]{1997ApJ...487..197E}
Ellison, D.~C., Drury, L.~O., \& Meyer, J.-P. 1997, \apj, 487, 197

\bibitem[{Epstein(1924)}]{Epstein:1924tc}
Epstein, P.~S. 1924, Physical Review, 23, 710

\bibitem[{Giacalone {et~al.}(2009)Giacalone, Giacalone, Jokipii, \&
  Jokipii}]{2009ApJ...701.1865G}
Giacalone, J., Giacalone, J., Jokipii, J.~R., \& Jokipii, J.~R. 2009, \apj,
  701, 1865

\bibitem[{Goldreich \& Scoville(1976)}]{1976ApJ...205..144G}
Goldreich, P., \& Scoville, N. 1976, \apj, 205, 144

\bibitem[{Guhathakurta \& Draine(1989)}]{1989ApJ...345..230G}
Guhathakurta, P., \& Draine, B.~T. 1989, \apj, 345, 230

\bibitem[{Hoang {et~al.}(2017)Hoang, Lazarian, Burkhart, \&
  Loeb}]{2017ApJ...837....5H}
Hoang, T., Lazarian, A., Burkhart, B., \& Loeb, A. 2017, \apj, 837, 5

\bibitem[{Hoang {et~al.}(2012)Hoang, Lazarian, \& Schlickeiser}]{Hoang:2012cx}
Hoang, T., Lazarian, A., \& Schlickeiser, R. 2012, \apj, 747, 54

\bibitem[{Hoang {et~al.}(2015)Hoang, Lazarian, \&
  Schlickeiser}]{2015ApJ...806..255H}
Hoang, T., Lazarian, A., \& Schlickeiser, R. 2015, \apj, 806, 255

\bibitem[{Hoang \& Loeb(2017)}]{Hoang:2017um}
Hoang, T., \& Loeb, A. 2017, arXiv:1706.07798

\bibitem[{Kipping(2017)}]{2017AJ....153..277K}
Kipping, D. 2017, \aj, 153, 277

\bibitem[{Kulkarni {et~al.}(2016)Kulkarni, Lubin, \&
  Zhang}]{2016SPIE.9981E..06K}
Kulkarni, N., Lubin, P.~M., \& Zhang, Q. 2016, in Proceedings of the SPIE,
  Univ. of California, Santa Barbara (United States), 998106

\bibitem[{Lee \& Cleaver(2016)}]{Lee:2016jb}
Lee, J.~S., \& Cleaver, G.~B. 2016, Modern Physics Letters A

\bibitem[{Linsky \& Wood(1996)}]{1996ApJ...463..254L}
Linsky, J.~L., \& Wood, B.~E. 1996, \apj, 463, 254

\bibitem[{Mathis {et~al.}(1977)Mathis, Rumpl, \& Nordsieck}]{Mathis:1977p3072}
Mathis, J.~S., Rumpl, W., \& Nordsieck, K.~H. 1977, \apj, 217, 425

\bibitem[{Mcinnes \& Brown(1990)}]{1990JSpRo..27...48M}
Mcinnes, C.~R., \& Brown, J.~C. 1990, \jspro, 27, 48

\bibitem[{Netzer \& Elitzur(1993)}]{1993ApJ...410..701N}
Netzer, N., \& Elitzur, M. 1993, \apj, 410, 701

\bibitem[{Noerdlinger(1971)}]{1971Ap&SS..13...70N}
Noerdlinger, P.~D. 1971, Ap\&SS, 13, 70

\bibitem[{Robertson(1937)}]{1937MNRAS..97..423R}
Robertson, H.~P. 1937, \mnras, 97, 423

\bibitem[{Scoville \& Norman(1995)}]{1995ApJ...451..510S}
Scoville, N., \& Norman, C. 1995, \apj, 451, 510

\bibitem[{Spitzer(1949)}]{Spitzer:1949bv}
Spitzer, L. 1949, Physical Review, 76, 583

\bibitem[{Wu {et~al.}(2015)Wu, Wang, Fan, \& et~al.}]{Wu:kr}
Wu, X.-B., Wang, F., Fan, X., \& et~al. 2015, Nature, 518, 512

\bibitem[{Yan \& Lazarian(2003)}]{2003ApJ...592L..33Y}
Yan, H., \& Lazarian, A. 2003, \apj, 592, L33

\bibitem[{Yan {et~al.}(2004)Yan, Lazarian, \& Draine}]{Yan:2004ko}
Yan, H., Lazarian, A., \& Draine, B.~T. 2004, \apj, 616, 895

\bibitem[{Ziegler {et~al.}(2010)Ziegler, Ziegler, \&
  Biersack}]{2010NIMPB.268.1818Z}
Ziegler, J.~F., Ziegler, M.~D., \& Biersack, J.~P. 2010, \nimpb, 268, 1818

\end{thebibliography}


\end{document}